\documentclass[trackchanges, twocolumn]{aastex7}

\usepackage{amsmath}
\usepackage{graphicx}
\usepackage{comment}
\usepackage[export]{adjustbox}
\begin{document}

\title{Size, Shape, and Geometric Albedo of Dwarf Planet Quaoar's Largest Satellite Weywot}

\author[0009-0004-7149-5212]{Cameron Collyer}
\affiliation{Florida Space Institute, University of Central Florida, 12354 Research Parkway, Orlando, FL 32826, USA}
\affiliation{Department of Physics, University of Central Florida, 4111 Libra Drive, Orlando, FL 32816, USA}
\email[show]{cpcollyerastro@gmail.com}  

\author[0009-0004-7149-5212]{Estela Fernández-Valenzuela}
\affiliation{Florida Space Institute, University of Central Florida, 12354 Research Parkway, Orlando, FL 32826, USA}
\email{}

\author[orcid=0000-0002-1788-870X,sname='Proudfoot']{Benjamin Proudfoot}
\affiliation{Florida Space Institute, University of Central Florida, 12354 Research Parkway, Orlando, FL 32826, USA}
\email{benp175@gmail.com}

\author[0000-0002-6117-0164]{Bryan J. Holler}
\affiliation{Space Telescope Science Institute, Baltimore, MD 21218, USA}
\email{bholler@stsci.edu}

\author[0000-0002-8690-2413]{Jose Luis Ortiz}
\affiliation{Instituto de Astrofísica de Andalucía, Glorieta de la Astronomía s/n, 18008 Granada, Spain}
\email{}

\author[0000-0002-8447-4857]{Ishan Sharma}
\affiliation{Department of Mechanical Engineering IIT Kanpur, Kanpur 208016, India}
\email{}

\begin{abstract}
    \noindent The small satellites of dwarf-planet size worlds in trans-Neptunian space are thought to be produced through giant impacts, and are typically observed to have bright water-ice-dominated surfaces. Here we focus on Quaoar's small primary satellite, Weywot. We build on previous work by re-analyzing the June 2023 stellar occultation by Weywot using results from 11 orbits of HST imaging with the Wide Field Camera 3. Weywot's rotational lightcurve was constrained along with its independent photometric phase curve, which we used to obtain more robust measurements of Weywot's size, shape, and geometric albedo. We found Weywot's volume-equivalent radius to be $r_{\rm eq} = 73 \pm 5$ km, with principal axes $a = 82^{+5}_{-3}$, $b = 73^{+5}_{-3}$, $c = 65^{+8}_{-12}$ km, indicating a triaxial shape. This shape is consistent with a Roche-ellipsoid rubble-pile interpretation requiring only moderate internal friction at Weywot's current semimajor axis, and Weywot can retain this shape down to Quaoar's co-rotation radius. We also found Weywot's geometric albedo to be quite low, $p_V = 0.036^{+0.010}_{-0.006}$, akin to that of Eris's satellite Dysnomia, but over an order of magnitude lower than Pluto's minor satellites and Haumea's satellites. We propose this may indicate that Weywot's ice-mass fraction $f_{\rm ice} < 100$\%, and that Weywot is not sourced from the water-ice mantle of a differentiated pre-impact progenitor as suggested for Pluto's minor satellites and Haumea's satellites. 
    
\end{abstract}

\section{Introduction}
Studies of trans-Neptunian binary/satellite systems (TNBs) provide critical insights into the formation and evolution of planetary building blocks. Their observations over the years have shed light on common planetesimal formation mechanisms \citep{Benecchi2009, Nesvorny2019streaming, Grundy2019}, provided constraints for the dynamical evolution of the outer solar system \citep{Parker2010Destruction}, and have given an indirect look into the subsurface structure of dwarf planets \citep{Nimmo2023, Regaly2025, Akiba2025}. Their rich diversity in orbital configurations and relative size between components represent their splaying evolutionary trajectories \citep{Noll2020Binaries, Bernstein2023}.

Among the most massive TNBs, there is substantial evidence that giant impacts played a role in satellite formation for at least two of them: Pluto and Haumea \citep{Canup2005Charon, Brown2007, Canup2011Irregular, Leinhardt2012Collisions}. Additional insight into evolutionary histories of two other TNBs can be gleaned from the ultra-slow rotation periods of Eris and Salacia, which imply outward migration of their satellites (Dysnomia and Actaea, respectively) over time \citep{Szakats2023, Bernstein2023, Collyer2025}. \cite{Arakawa2019} presented simulations of giant impacts using a range of collisional orientations and progenitors, and showed that satellite systems of all the largest TNBs are comprehensively reproduced through these collisions. 

Weywot and the newly discovered undesignated moon orbit Quaoar beyond its two rings \citep{Nolthenius2025Moonlet, Morgado2023, Pereira2023}. Weywot's small inclination relative to the ring-plane \cite[$\lesssim 5^{\circ}$,][]{BragaRibas2025,Proudfoot2025Weywot} and prograde orbit support its formation from a circum-Quaoar disk. Quaoar's multi-component system may be sourced from the debris disk of a giant impact early in its history.

Weywot is among the class of small satellites around the largest trans-Neptunian objects (TNOs), which we define here as having secondary-to-primary mass ratios $q \lesssim 0.5$\%. This class includes Haumea's satellites, Pluto's minor satellites, Makemake's satellite S/2015 (136472) 1, Eris's satellite Dysnomia, and Quaoar's recently discovered second moon \citep{Proudfoot2025haumea, Porter2023MassesIrregular, Parker2016mk2, Brown2023, Nolthenius2025Moonlet}. The bright, water-ice dominated surfaces of Haumea's satellites and Pluto's minor satellites have been reproduced from collisions between fully or partially differentiated progenitors, where the satellites are sourced from the water-rich mantles of the progenitors \citep{Fraser2009HaumeaSatellites, Canup2011Irregular, Cook2018Irregular, Arakawa2019, Estela2025Hiiaka}. In these simulations, satellites with $q \lesssim 0.5$\% have ice-mass fractions of nearly 100\%, independent of their formation from a debris disk or as a collisional fragment, which is consistent with their bright surfaces. Dysnomia's surface, on the other hand, is quite dark \citep{Brown2023}, and may not fit this paradigm. The surfaces of S/2015 (136472) 1 and Quaoar's newly discovered moon are currently unknown. 

In this study, we build on the \cite{Estela2026} work on Weywot by incorporating 11 orbits of HST imaging with the Wide Field Camera 3 (WFC3). Specifically, we constrain Weywot's rotational lightcurve, identify its lightcurve phase during the June 2023 occultation, and measure Weywot's independent photometric phase curve for more robust constraints of its absolute magnitude, size, 3D shape and geometric albedo. Our results place Weywot among the most well-defined satellites in trans-Neptunian space, and calls for future targeted investigations of its precise surface chemistry.


\section{Observations, Reduction, and Photometry}

\begin{deluxetable}{ccccccccc}
\tablecaption{Photometry \label{obstable}} 
\tablehead{ & & \multicolumn{2}{c}{Quaoar} & \multicolumn{2}{c}{Weywot} & & \\    Date & Filter & $m_{Q}$ & $\sigma_{m_{Q}}$ & $m_{W}$ & $\sigma_{m_{W}}$ & $r$ & $\Delta$ & $\alpha$\\ (MJD) & & (mag)& (mag)& (mag)& (mag)& (au) & (au) & ($^{\circ}$)}
\startdata
60451.526650 & F350LP & 19.0954 & 0.0097 & 24.642 & 0.022 & 42.681 & 41.893 & 0.866 \\
60483.397265 & F350LP & 18.8482 & 0.0096 & 24.473 & 0.022 & 42.678 & 41.680 & 0.267\\
60760.661680 & F350LP & 19.0883 & 0.0050 & & &              42.652 & 42.724 & 1.337 \\
60763.744867 & F350LP & 19.2276 & 0.0077 & 24.890 & 0.018 & 42.652 & 42.671 & 1.343\\
60769.979372 & F350LP & 19.2038 & 0.0036 & 24.791 & 0.023 & 42.651 & 42.565 & 1.343\\
60856.540736 & F606W & 18.9844 & 0.0094 & 24.450 & 0.027 &  42.643 & 41.637 & 0.198\\
60858.109059 & F606W & 18.9823 & 0.0077 & 24.552 & 0.025 &  42.643 & 41.637 & 0.195\\
60860.008315 & F606W & 18.9254 & 0.0036 & 24.544 & 0.030 &  42.643 & 41.637 & 0.200\\
60861.517938 & F606W & 18.876 & 0.011 & 24.453 & 0.024 &    42.643 & 41.638 & 0.211\\
60863.086307 & F606W & 18.883 & 0.012 & 24.544 & 0.029 &    42.643 & 41.640 & 0.227\\
60864.910282 & F606W & 18.845 & 0.010 &  &  &               42.643 & 41.643 & 0.251\\
\enddata
\tablecomments{Each MJD date is the mid-time of all exposures in that orbit, corrected for light travel time.}
\end{deluxetable}

\subsection{Hubble Space Telescope Observations}
\label{sec:observations}
HST observations analyzed in this work spanned May 21, 2024, through July 8, 2025, and were taken with the WFC3 instrument and the UVIS2 channel. Data from late May 2024 through early April 2025 were part of the HST Cycle 31 program 17417 \citep[PI: B. Proudfoot, see][]{Proudfoot2025Weywot}, where the F350LP longpass filter was used (hereafter, F350LP data). The F350LP data were taken over 5 orbits, with eight frames per-orbit, using a four-point dither pattern. Exposure times per-frame varied from 220 s to 240 s. Data from late June through early July 2025 were part of the HST Cycle 32 program 17827 (PI: E. Fernandez-Valenzuela) where the F606W filter was used (hereafter, F606W data). F606W data was taken over seven orbits, with four frames per-orbit, using a four-point dither pattern. The seventh orbit failed due to failed guide star acquisition, so data was measured only on the first six orbits. All F606W exposures were 352 s. The telescope tracked at the rate of motion of the Quaoar-Weywot system, causing background stars to trail in the images. We show the median-combined fully calibrated images from each orbit in Figure \ref{figure:photometry}.

\subsection{PSF Photometry Methods and Results}
\label{sec:photometry}
We measured photometry from the fully calibrated, flat-fielded, and charge transfer efficiency-corrected frames provided by the Mikulski Archive for Space Telescopes (MAST). We used PSF photometry to extract the flux of Quaoar and Weywot in images from both filters using empirical PSFs provided by the Space Telescope Science Institute (STScI) \citep{Dauphin2021PSFs}.

\subsubsection{F350LP Images}
The F350LP filter spans approximately 0.350 $\mu$m - 1.05 $\mu$m. Over such a large wavelength range, the resulting PSF shape of a source on the detector is spectrum-dependent. As such, STScI does not provide empirical PSFs for this filter. We created an empirical PSF using the reflectance spectrum of Quaoar and the standard empirical PSFs for the F390W, F606W, and F814W filters. Over the 0.350 $\mu$m $- 1.05$ $\mu$m wavelength range, we estimated Quaoar's albedo spectrum as linear with a spectral slope $\mathcal{S}$ = 28 \%/100 nm, normalized by the reflectance at $\lambda = 0.65$ $\mu$m \citep{JewittQuaoar}. Combining Quaoar's spectrum with the solar spectrum provided by the \texttt{sbpy} Python package \citep{sbpy}, we estimated the flux contribution for each filter by convolving each filter's throughput\footnote{\url{https://www.stsci.edu/hst/instrumentation/wfc3/performance/throughputs}} with the model spectrum and numerically integrating the flux density over the wavelength range. The empirical PSFs were then scaled by their relative contributions, added linearly, and normalized. 

The PSF photometry pipeline was developed and improved upon in multiple steps. The Python notebooks provided by STScI\footnote{\url{https://spacetelescope.github.io/hst_notebooks/notebooks/WFC3/point_spread_function/hst_point_spread_function.html}} were extremely helpful for reference. At the earliest stage, we estimated the flux from Quaoar and Weywot in individual frames by doing a double PSF fit. Since Weywot is only $\sim0.6$ \% of Quaoar's flux, separating Quaoar's flux contribution from that of Weywot's is crucial for accurate photometry of Weywot. 

\begin{figure*}
\centering
 	\includegraphics[scale = 1.2, valign = t]{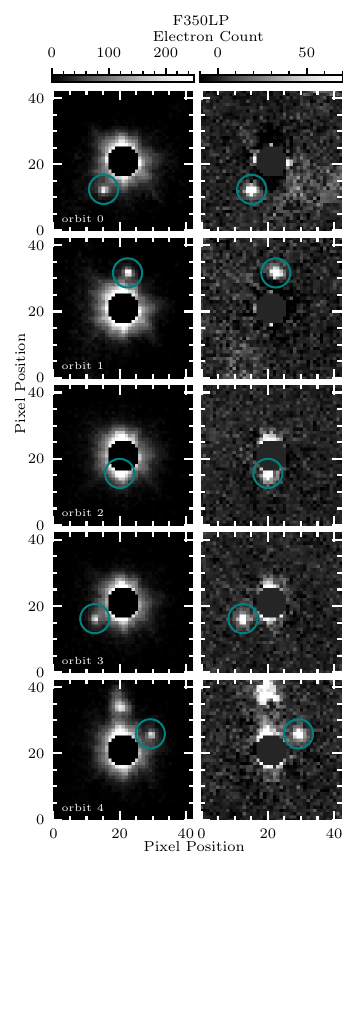}
    \includegraphics[scale = 1.2, valign = t]{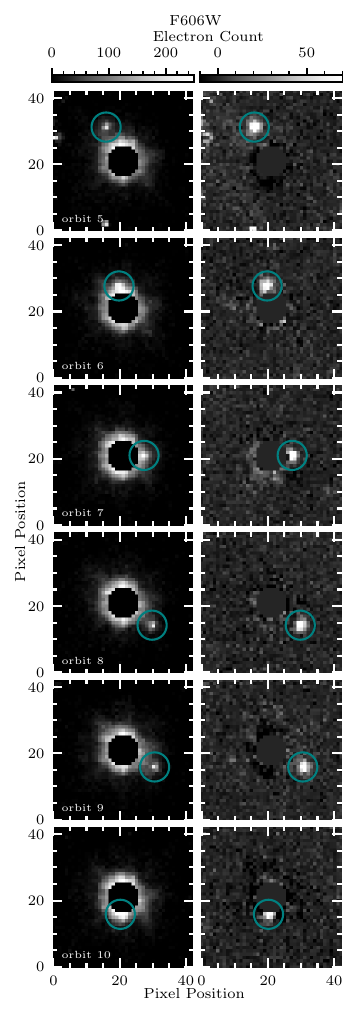}
	\caption{F350LP and F606W images before and after Quaoar subtraction. Quaoar is masked in all images to enhance Weywot's visibility. Weywot's position is circled. Due to its close proximity to Quaoar in orbit numbers two and ten, photometry of Weywot from these orbits were not trusted (see Section \ref{popdrop}).} 
 \label{figure:photometry}
\end{figure*}

In each frame post-PSF subtraction, there was a clear asymmetric residual flux structure due to Quaoar, indicating the model PSF does not provide a fully realistic rendering of the WFC3 PSF. This was in part due to Quaoar not being a true point source, but was also the result of the increased jitter in HST's pointing stability due to the transition to single-gyro mode \citep[e.g.,][]{AndersonJitter}. We found that the PSF symmetry degraded over the course of an orbit. To remove the residual flux from Quaoar and its evolving structure over the course of an orbit, residual background frames were created from Quaoar-removed frames where no contaminating sources were nearby. Combining the residual frames normalized by Quaoar's flux from each orbit in groups of two (frames 1\&2, ..., frames 7\&8), four master residual background frames were generated by median-combining residual frames from each of the 5 orbits. Since Weywot's position relative to Quaoar changed each orbit, this process allowed us to sample Quaoar's residual flux structure at Weywot's location in each frame. The residual background frame for a given image was scaled to the flux of Quaoar in that image and subtracted. This process flattened the resulting background considerably, isolating Weywot's signal. We emphasize that this residual background subtraction substantially improved the resulting photometry of Weywot. In Figure \ref{figure:photometry} we show the resulting F350LP frames on the left with Quaoar's contribution removed in the second column. Note that the elevated background levels seen in the first two orbits is due to multiple sources slowly crossing those regions over the course of the eight frames. 

\subsubsection{F606W Images}
The F606W images were calibrated following similar methods applied to the F350LP data. Since four F606W images were taken per-orbit as opposed to the eight F350LP images per-orbit, master residual background frames were created for each of the four dither positions. Again, Weywot moved with respect to Quaoar over the six orbits where the F606W data were taken, allowing the residual flux structure from Quaoar to be properly sampled at the location of Weywot in each orbit. This process flattened the resulting background substantially. In Figure \ref{figure:photometry} we show the resulting F606W frames on the right with Quaoar's contribution removed in the second column.

\subsubsection{Pop-and-Drop PSF Re-sampling of Weywot}
\label{popdrop}
Initial lighturve analysis of preliminary photometry showed that Weywot's flux decreased in images where it was closer to Quaoar. While this is expected if Weywot is elongated and its rotation rate is synchronized with its orbit, we ran additional tests to ensure that Weywot's flux drops were not due to over-subtraction of Quaoar's flux when it was in closer proximity. In particular, in orbits two, six, seven and ten, Weywot was very close to the region where inadequacies of the model PSF can strongly affect Weywot's photometry (central area masked in Figure \ref{figure:photometry}). 

In order to test whether Weywot's measured flux variations were only due to imperfect PSF subtraction, we ran the following tests, which we refer to as the `Pop-and-Drop' method of PSF resampling. We discuss the method applied to the F606W data as an example. 

The location of Weywot with respect to Quaoar was measured in every frame across the six orbits. For each Quaoar-removed frame in an orbit, the PSF was fitted to Weywot and a flux was measured, allowing the creation of a Weywot model PSF. This model PSF was then subtracted from the image (Pop), and iteratively shifted and placed at the location of Weywot in every other Quaoar-removed image (Drop), barring the other images from that orbit. The flux of Weywot was then measured at the other positions, effectively resampling the flux across the residual frame. This allowed us to determine in which regions our PSF photometry methods could be trusted. In the F606W data, we found that the resampled fluxes at the location of Weywot in orbit 10 were extremely discrepant. At all other positions the resampled fluxes followed a Gaussian distribution around the injected fluxes. In the F350LP data, we found that the recovered fluxes at Weywot's position in orbit 2 was also discrepant. Unsurprisingly, both of these orbits had approximately the same position on the WFC3 detector (relative to Quaoar).

\subsubsection{Photometry Results}
The flux of Quaoar measured in each orbit is the median of the flux across the respective frames, and its error is the standard deviation on the mean. We report the resulting apparent magnitudes on the VEGAMAG scale in Table \ref{obstable}, where the photometric zeropoints of $Z_{\text{F350LP}}$ = 26.780 mag and $Z_{\text{F606W}}$ = 25.995 mag were applied\footnote{\url{https://www.stsci.edu/hst/instrumentation/wfc3/data-analysis/photometric-calibration/uvis-photometric-calibration}}.

The flux of Weywot in each orbit was also measured as the median of the flux across the frames. We used a slightly different method to estimate the uncertainty on the flux of Weywot. For each image $i$ in a given orbit of length $N$ images, we used Pop-and-Drop resampling (Section \ref{popdrop}) to resample the flux at the other Weywot locations (barring the fluxes at the locations of Weywot in orbits 2 and 10) and took the median of these resampled fluxes ($F_{r,i}$). The standard deviation on the mean of the set of fluxes [$F_{r,1}$, ..., $F_{r,N}$] was taken as the uncertainty on the orbit flux of Weywot. We used this method since it yielded uncertainties that were always larger or comparable to the standard deviation on the mean of the fluxes over the orbit.

\section{Lightcurve Model and Assumptions}
\subsection{Model Assumptions}
\label{assumptions}

In constraining Weywot's rotational lightcurve from the resolved HST observations, we make the foundational assumption that Weywot's rotation period is synchronized with its orbit. We also assume Weywot's rotation pole is aligned with the mutual orbit pole when deriving Weywot's shape model from the June 2023 occultation and lightcurve parameters. Here, we provide estimates of the timescales for spin synchronization and obliquity damping for Weywot. Synchronous rotation and zero obliquity is the equilibrium state of Weywot's tidal evolution. Although there is a nonzero probability of Weywot being trapped in a higher order spin-orbit resonance if its shape is elongated and the orbit is eccentric, this was also true for many of the satellites in the Solar System which passed these resonances and now exhibit synchronous rotation. Additionally, \cite{Gladman1996} showed that small satellites around larger primaries evolve towards the first Cassini state --- equivalent to zero obliquity in the case of Weywot. Estimates for the timescales of spin-synchronization \citep{Goldreich1966} and obliquity damping \citep{Boue2019Tides}, assuming the eccentricity and initial obliquity are small, follow:

\begin{equation}
    \tau _{\text{spin}} = \left| \frac{1}{P_f} - \frac{1}{P_0} \right| \left( \frac{45 \pi G}{4} \frac{R_s ^3 M_p^2}{m_s} \frac{1}{a^6} \frac{1}{Q'}    \right)^{-1}
\end{equation}

\begin{equation}
\tau_{\text{obliq}} =  \text{ln}\left(\frac{i_{0}}{i_{f}}\right) \left( \frac{9}{4}n(1 + \varrho_s) \frac{M_p}{m_s} \left (  \frac{R_s}{a} \right)^5\frac{1}{Q'}\right)^{-1},
\end{equation}

\noindent respectively, where:

\begin{equation}
    Q' = Q\left(1 + \frac{57}{8 \pi G} \frac{\mu}{R_s^2 \rho_s^2}\right),
\end{equation}

\begin{equation}
    \varrho_s = \frac{M_p m_s}{M_p + m_s}\frac{a^2}{I_s}.
\end{equation}

\noindent In the above equations, $P_f$ and $P_0$ refer to the final and initial spin periods, $G$ is the gravitational constant, $R_s$, $m_s$, and $\rho_s$ refer to Weywot's radius, mass and density, respectively, $a$ is the current semimajor axis, $\mu$ is the shear modulus, $Q$ is the dimensionless tidal dissipation (quality) factor, and $I_s = (2/5) m_s R_s ^2$ is Weywot's moment of inertia. Canonical values of $\mu$ and $Q$ for icy bodies are $\mu = 4$ GPa \citep{Neumeier2018} and $Q=100$ \citep{Murray2000}. Quaoar's mass was estimated in \cite{Proudfoot2025Weywot} to be $M_Q = (1212.27 \pm 5.39) \times 10^{18}$ kg and \cite{Estela2026} found that Weywot's radius lies between $R_W = 58-86$ km. Note that the timescales are insensitive to uncertainties in Quaoar's mass, Weywot's semi-major axis, and Weywot's radius. 

The value of $\tau_{\text{spin}}$ increases when the initial spin period is fast, Weywot's density is low, $Q$ is large, and $\mu$ is large. Assuming canonical values for the tidal parameters, $\mu = 4$ GPa and $Q=100$, a minimum Weywot density of $400$ kg m$^{-3}$ and Weywot's breakup spin period $P_0 = 2.5$ hr, $\tau _{\text{spin}} = 150$ Myr. If Weywot were a rubble pile, then the shear modulus would be reduced and so would $\tau _{\text{spin}}$. This implies that it is safe to assume Weywot's spin to be synchronized with the orbit period of $P_f \simeq 12.43$ days.

The value of $\tau_{\text{obliq}}$ increases with initial obliquity, $Q$ and $\mu$, and decreases with increasing density. Assuming canonical values for the tidal parameters, $\mu = 4$ GPa and $Q=100$, a minimum Weywot density of $400$ kg m$^{-3}$, an initial and final obliquity of $i_0=25^{\circ}$ and $i_f=0.5^{\circ}$, $\tau_{\text{obliq}} = 200$ Myr. If Weywot were a rubble pile the shear modulus would be reduced and so would $\tau _{\text{obliq}}$. Again, such a short timescale implies our assumption that Weywot's obliquity has been fully damped is safe.

These computed timescales are strong upper limits based on the current orbit of Weywot. If Weywot has migrated outward while spinning down Quaoar, these timescales would be significantly shorter due to their strong inverse dependence on semimajor axis.

\subsection{Weywot's Synchronized Rotation Period}
\begin{figure}
\centering
 	\includegraphics[scale = 0.70]{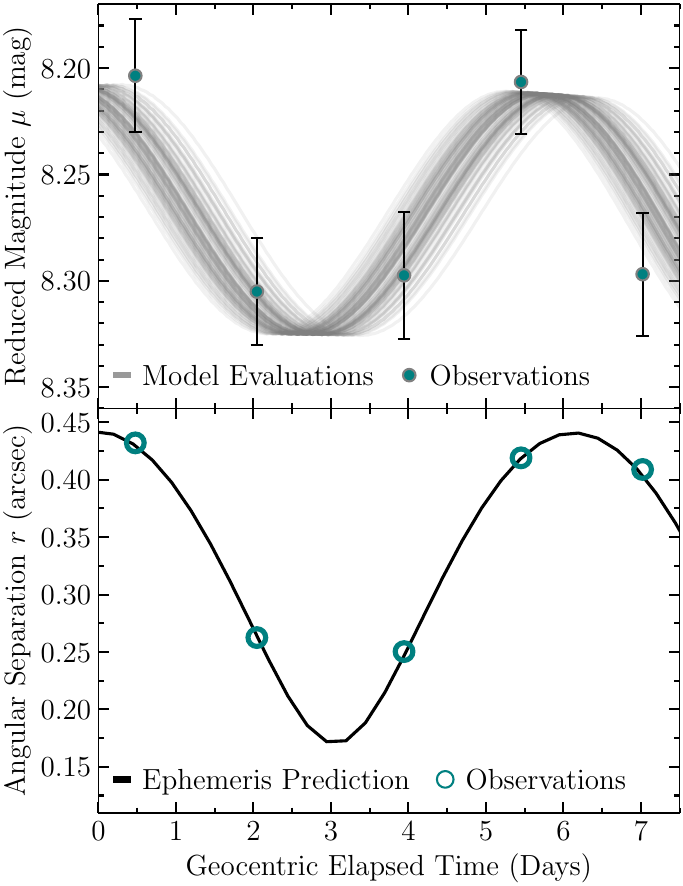}
	\caption{
    \textbf{Top}: F606W-band photometry. These data were taken consecutively with a spacing of $\approx 1.5$ days. The $x$-axis is in geocentric elapsed time, where $t=0$ coincides with an epoch of maximum elongation of Weywot from Quaoar. The 100 model evaluations correspond to the lightcurve solution presented below (Eq. \ref{eq:fullmodel}), allowing $\phi$ to vary within its uncertainty. \textbf{Bottom}: Angular separation of Weywot from Quaoar versus time. The black line is the prediction based on the published ephemeris. The fact that the photometry and angular separation come to maximum and minimum at similar times supports Weywot's spin being synchronized to the orbit, and also indicates a triaxial shape for Weywot.}
    \label{figure:m_r}
\end{figure}

We show in Figure \ref{figure:m_r} additional evidence that supports the assumption that Weywot's spin is synchronized with its orbit. The $x$-axis is in geocentric elapsed time where $t=0$ coincides with an epoch of maximum elongation between Quaoar and Weywot. The two panels only incorporate F606W data. In the top panel is the F606W-band photometry corrected for geocentric and heliocentric distances. The grey model evaluations correspond to the lightcurve solution found below (Eq. \ref{eq:fullmodel}), which incorporates the F606W and F350LP data, allowing $\phi$ to vary within its uncertainty. In the bottom panel we show the radial angular separation between Quaoar and Weywot, predicted from the \cite{Proudfoot2025Weywot} ephemeris (black line) and measured from the images (teal rings) \footnote{Astrometric Measurements from the F606W Images are provided in Appendix \ref{astrometry}.}.

This figure indicates that the the photometry oscillates in phase with the angular separation, exactly what would be expected if Weywot's shape is triaxial and its spin is synchronized with its orbit. A slight timing offset between the model lightcurve and angular separation extrema is likely the result of an imperfect model (see Section \ref{lcmodel}), and not a real phase-lag. We therefore assume Weywot's sidereal rotation period to be 12.427 days, and we demonstrate in Appendix \ref{synodic} that this can be taken to be the perceived synodic rotation period during observations.

\subsection{Lightcurve Model}
\label{lcmodel}


Our model for Weywot's rotational lightcurve must account for the changing phase angle over the course of the observations ($0.19 ^{\circ}  \lesssim \alpha \lesssim 1.34^{\circ}$, Table \ref{obstable}). Over such a small range in phase angle, small to midsize TNOs exhibit a linear relationship where apparent magnitude increases with phase angle \citep[e.g.,][]{Rabinowitz2007}. Weywot, a relatively low albedo satellite \citep{Estela2026} spectro-photometrically characterized as an H$_2$O-type (A. C. Souza Feliciano, submitted) is expected to have a mature surface, and therefore we choose a linear model for its phase behavior \citep[e.g.,][]{Verbiscer2022}. We assume an identical phase curve for Weywot in the F606W and F350LP filters. While the phase behavior can be filter-dependent \citep[e.g.,][]{Rabinowitz2007}, this should have minimal effect on our results; during the $\sim 8$ days over which the F606W data were taken, the phase angle changed by only $\Delta \alpha \simeq0.05^{\circ}$. In light of the results presented in Section \ref{lcresults}, we find that the data do not show evidence of deviation from a linear trend between magnitude and phase angle.

The apparent magnitudes ($m$) from Table \ref{obstable} were transformed to \textit{reduced} magnitudes ($\mu$), the magnitude transformed to a heliocentric ($r$) and geocentric ($\Delta$) distance of 1 au:

\begin{equation}
    \mu = m - 5 \text{log} \left(r\Delta\right),
\end{equation}

\noindent where $r$ and $\Delta$ are input in units of au. The rotational-phase dependent $absolute$ magnitude $h$ is the reduced magnitude transformed to a phase angle of $0^{\circ}$. After converting to absolute magnitude, any residual magnitude fluctuations seen from Weywot are due to its rotational lightcurve. Inputting $\alpha$ in units of degrees, with $\beta$ being the slope with units of mag deg$^{-1}$,

\begin{equation}
\label{phasecurve}
    h = \mu - \beta \alpha.
\end{equation}

For Weywot, we use a simple sinusoid to model the lightcurve due to the small number of datapoints ($n=9$)

\begin{equation}
\label{eq:simple}
    h = H +\left( \frac{\Delta m}{2} \right) \text{cos} \left(  4\pi \left(\frac{t}{P}\right) + 2\phi \right),
\end{equation}

\noindent where we have defined $H$ as the absolute magnitude averaged over a rotation. In the above equation, we assume the lightcurve shows two oscillations over the rotation period ($P$), which is accurate for a triaxial body and supported by Figure \ref{figure:m_r}. $\Delta m$ is the peak-to-peak amplitude, and $\phi$ is the phase-offset. We define $t=0$ as the mid-time of the June 2023 occultation of Weywot (see Section \ref{occ}), thus $\phi$ gives the lightcurve phase of Weywot during the occultation. 

Equating the above two equations, keeping in mind Weywot's synchronized rotation period, our chosen model for the observational data $\mu$, $\alpha$ and $t$ is

\begin{equation}
\label{eq:fullmodel}
    \mu = H +\left( \frac{\Delta m}{2} \right) \text{cos} \left(  4\pi \left(\frac{t}{P}\right) + 2\phi \right) + \beta \alpha.
\end{equation}

To combine the F606W and F350LP datasets into a composite lightcurve, we must also incorporate a color offset between the two filters. In doing so we assume the lightcurve shape and amplitude for both filters is equivalent. We define the color offset 

\begin{equation}
    \mathcal{C} = H_{\text{F606W}} - H_{\text{F350LP}},
\end{equation}

\noindent where we convert F606W-band data to F350LP-band using the fit parameter $\mathcal{C}$. Thus, our preferred model of the observational data $\mu$, $\alpha$ and $t$ has five fit parameters: $H$, $\beta$, $\mathcal{C}$, $\Delta m$, and $\phi$.

We introduce a second lightcurve model with an additional two fit parameters:

\begin{equation}
\label{eq:fourier}
\begin{split}
    h = H +  A \ \text{cos} \left(2\pi \left( \frac{t}{P} \right) + \phi_1\right) + \\
    \left( \frac{\Delta m}{2} \right) \text{cos} \left(4\pi \left( \frac{t}{P} \right) + 2\phi_2\right).
\end{split}
\end{equation}

\noindent Therefore, our two lightcurve models are Eqs. \ref{eq:simple} and \ref{eq:fourier}, where Eq. \ref{eq:simple} is nested in Eq. \ref{eq:fourier}. With respect to Weywot, under the assumption of synchronous rotation, we show below that the photometry points to two nearly symmetric oscillations over a single synchronous period. This indicates that Weywot's 3D shape is prolate \citep[e.g.][Figure 1]{Thirouin2014}. The additional term in Eq. \ref{eq:fourier} has a single minimum and maximum over the period $P$, introducing an asymmetry in the resulting lightcurve. Thus, we relate the amplitude in Eq. \ref{eq:simple} to Weywot's shape, and refer to the amplitude ($A$) in Eq. \ref{eq:fourier} as the \textit{asymmetry} parameter. While these two models likely do not capture the full complexity of Weywot's true lightcurve shape, they can provide a blueprint for Weywot's overall shape (Eq. \ref{eq:simple}) and indicate a nonuniform surface morphology (Eq. \ref{eq:fourier}), e.g., Salacia \citep{Collyer2025}. 

\section{Results}
\subsection{Weywot's Rotational Lightcurve}
\label{lcresults}

\begin{figure*}
\centering
 	\includegraphics[scale = 0.70]{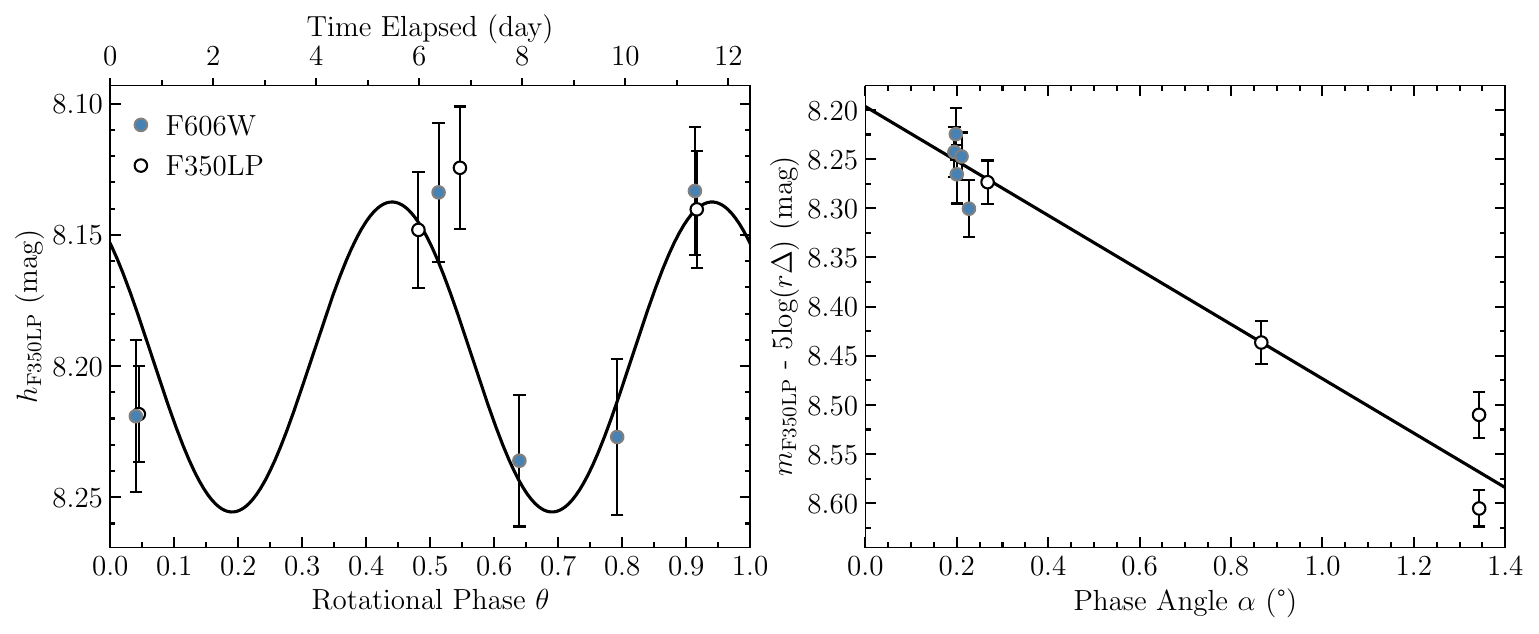}
	\caption{
    MCMC solution for Weywot's rotational lightcurve (left) and photometric phase-curve (right). The rotational lightcurve is assumed to be a simple sinusoid, and a linear relationship is assumed between reduced magnitude and phase angle. Table \ref{W_lcparams} contains the solution for the model parameters. $\theta = t/P = 0$ corresponds to the rotational phase during the occultation.
 \label{figure:Weywot_simple_lc}
 }
\end{figure*}

With the photometric and lightcurve model developed above, we employ a Bayesian approach to determine solutions consistent with the observed photometry of Weywot using the \texttt{emcee} \citep{foreman2013emcee} Markov chain Monte Carlo (MCMC) sampler. We first explored solutions for the single-amplitude (symmetric) lightcurve model (Eq. \ref{eq:simple}), using uniform priors on each parameter. The recovered lightcurve and phase curve are shown in Figure \ref{figure:Weywot_simple_lc} and the posterior results for the fit parameters $H$, $\beta$, $\mathcal{C}$, $\Delta m$, and $\phi$ are tabulated in Table \ref{W_lcparams} \footnote{See Appendix \ref{cornerplots} for corner plots.}. We see in Figure \ref{figure:Weywot_simple_lc} that the resulting photometric and lightcurve model fit the observational data well, with $\chi^2_{\text{pdf}} = 3.42$. Moreover, at rotational phases ($\theta$) where F606W and F350LP data nearly overlap, the converted photometry is in close agreement. Indeed, at $\theta \simeq 0.05$ the fact that the F350LP and F606W data produce a nearly identical residual to the model indicates that Weywot's true lightcurve is more complex than this simple model is able to reproduce. However, with $n=9$ datapoints, there are only $\nu = 4$ degrees of freedom in the fit, restricting our use of a more complicated model for the lightcurve. Implementing Eq. \ref{eq:fourier} as the lightcurve model, allowing asymmetric minima and maxima and reducing the degrees of freedom to $\nu = 2$, results in an overfit, unconstrained model due to the lack of phase coverage between $0.05 \lesssim \theta \lesssim 0.5$. If we place a fine-tuned Gaussian prior on the asymmetry parameter with a mean $A = 0$ and width $\sigma_{A} = 0.01$, we obtain a more constrained fit, resulting in $\chi^2_{\text{pdf}} = 0.77$, still suggestive of overfitting. We take this as initial evidence that Weywot's lightcurve might be slightly asymmetric. Our preferred solution is based on the symmetric lightcurve, with parameters tabulated in Table \ref{W_lcparams}. 

In summary, the rotational lightcurve supports Weywot's spin period being synchronized to its orbit. The double-peaked lightcurve over one full rotation implies Weywot's 3D shape is prolate, though the small lightcurve amplitude measured at an aspect angle of $\psi \simeq 69^{\circ}$ rules out Weywot being a contact binary. The derived lightcurve parameters are used in Section \ref{occ} to derive Weywot's 3D shape.

\begin{deluxetable}{llcc}
\tablecaption{Weywot's Photometric and Lightcurve Parameters\label{W_lcparams}}
\tablewidth{0pt}
\tablehead{Parameter & Unit & Symbol & Posterior}
\startdata
F350LP Absolute Magnitude        & mag             & $H_{F350LP}$  & $8.197^{+0.029}_{-0.029}$  \\
Phase Coefficient                & mag deg$^{-1}$  & $\beta$       & $0.277^{+0.027}_{-0.027}$  \\
$H_{F606W}$ - $H_{F350LP}$ Color & mag             & $\mathcal{C}$ & $0.015^{+0.029}_{-0.029} $ \\
Lightcurve Amplitude             & mag            & $\Delta m$    & $0.119^{+0.036}_{-0.036} $ \\
Phase Offset                     & deg            & $\phi$        & $111.4 ^{+7.6}_{-6.7} $    \\
\hline
F606W  Absolute Magnitude        & mag             & $H_{F606W}$  & $8.211^{+0.041}_{-0.041}$  \\
$V$      Absolute Magnitude        & mag           & $H_{V}$      & $8.484^{+0.042}_{-0.042}$                        \\
\enddata
\tablecomments{Reported values show the median, 16th, and 84th percentile values. Measurements below the horizontal line are derived parameters. Conversion to $H_V$ is described in Appendix \ref{filterconversion}.}
\end{deluxetable}


\subsection{Occultation Analysis}
\label{occ}
Recently, a stellar occultation of Weywot enabled measurement of its approximate size and albedo \citep{Estela2026}. With our determination of Weywot's rotational lightcurve, further analysis of this event can give insight into Weywot's three-dimensional shape. To do this, our analysis uses \texttt{occult3d} \citep{Proudfoot2026Triaxial}, which can derive triaxial ellipsoid shape models by combining occultations, lightcurves, and other external constraints. It uses a Bayesian approach to explore triaxial shape models consistent with any combination of these data sets, specifically using the \texttt{emcee} MCMC sampler \citep{foreman2013emcee}. 

To derive a shape model, we first had to constrain Weywot's rotation axis, its lightcurve phase during the occultation, and its lightcurve amplitude. We found in Section \ref{assumptions} that the timescale for a nonzero initial obliquity of Weywot to be fully damped to be $\tau_{\text{obliq}} \leq 200$ Myr, implying Weywot's rotation pole is parallel to the well-constrained mutual orbit pole. Propagating forward the \citet{Proudfoot2025Weywot} orbit solution for Weywot to the time of the June 2023 occultation (including the precession of the orbit), we find that Weywot's rotation pole is $(\alpha_p, \delta_p) = (267.6\pm0.2 \degr,52.16\pm0.2 \degr)$, corresponding to an aspect angle $\psi \simeq 69 \degr$. The lightcurve amplitude and rotational phase during the occultation were measured in Table \ref{W_lcparams}. To incorporate these measurements into a shape model, \texttt{occult3d} uses them as priors, implemented as Gaussians within the code. 

The shape model implemented within \texttt{occult3d} has 8 free parameters, the ellipsoid semi-axes ($a$, $b$, $c$), the rotation pole direction ($\alpha_p, \delta_p$), the rotational phase ($\phi$), and the on-sky coordinates of Weywot at the time of the occultation ($f$, $g$), referenced to the \citet{Proudfoot2025Weywot} and NIMAv20 \citep{Desmars2015} ephemerides of Weywot and Quaoar, respectively. We fit our shape model to the occultation chords as reported in \citet{Estela2026}, although we did not use the RECON24 chord which was significantly discrepant with a more precise chord at a similar cross-track location. The most precise chord, observed from the 4.3-m Lowell Discovery Telescope (LDT), had timing uncertainties of 0.002 s, translating to $\sim$50 meter uncertainty on the chord extremity. The tiny uncertainty can provide significant barriers to high-quality global shape models as any departures from triaxiality---or equivalently any topographic features---will bias the global shape model. To solve this, we added in quadrature 5 km---approximately the size of expected topographic features \citep{Johnson1973}---to the chord extremity uncertainties (determined by the timing uncertainty) to account for any topography \citep[see similar implementations in][]{Morgado2021,SORAcitation,Rommel2023}. 

Although using \texttt{occult3d} alone can constrain Weywot's three-dimensional shape, the June 2023 occultation's sampling of Weywot's northern hemisphere was lacking \citep[as seen in Figure 4 of][]{Estela2026}, resulting in large uncertainties on Weywot's $c$-axis. To further constrain the shape model, we look to other reports of occultations by Weywot. \citet{BragaRibas2025} reported preliminary results of three additional single- or double-chord occultations from 2019--2023. Interestingly, these events provided chords of $168\pm15$ km (2019-08-04), $\sim135$ km (2022-06-11), and $180\pm19$ km (2023-05-26). Unfortunately, as neither observing station locations nor immersion/emersion timings were provided, we could not directly fit these occultation chords. However, with the chord lengths alone, we were able to incorporate some of their constraints into our models. 

To do this, we propagated each of our trial shape models---contained as samples within our MCMC chains---to the epoch of those occultations, accounting for both changes in viewing geometry and the rotational phase of Weywot. The rotational phase during each occultation was determined based on the orbital phase at the time of the occultation, consistent with Weywot's tidally locked state (see Figure \ref{figure:m_r}). For each MCMC sample, we then determined the elliptical limb profile during the single-/double-chord events. Based on the limb's major axis---the longest possible occultation chord---we then calculated the probability that the reported chords could fit within the elliptical limb during each occultation. 

\begin{figure}
    \centering
     \includegraphics[width=\columnwidth]{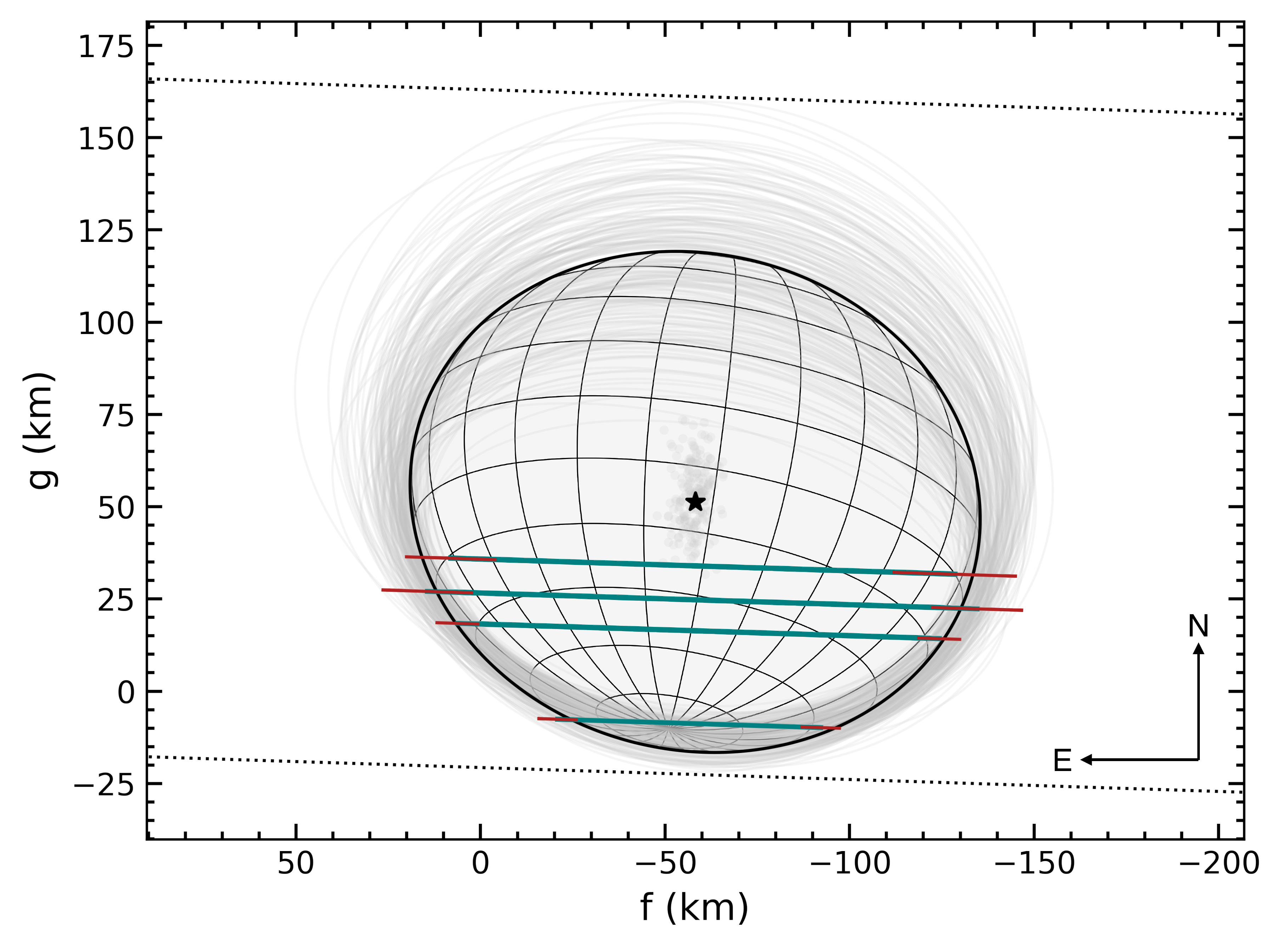}
    \caption{Limb profile and shape model for the June 2023 occultation. Blue lines show occultation chords, with chord extremity uncertainties in red (including 5 km uncertainty for topography). The solid black line and triaxial shape model show our best-fit solution. Gray lines show a sample of other possible limb profiles. Dotted lines show close negatives. }
    \label{fig:shapemodel}
\end{figure}

Since we had no further information about the chords, this process neglected any constraints given by the shadow path geometry, although we emphasize that the additional information will always be more constraining than using the chord length alone. We then removed all MCMC samples which had a combined probability of matching the three occultations of $<1\%$. This process eliminated $\sim70\%$ of the shape model parameter space, significantly shrinking the uncertainty in Weywot's shape, especially the $c$-axis, and only increased the June 2023 occultation's best-fit $\chi^2$ by 0.6. The best fit shape model provides excellent agreement with the reported chords during the June 2023 occultation (Figure \ref{fig:shapemodel}). 

Figure \ref{fig:shapemodel} shows the three-dimensional shape model overlaid with the June 2023 occultation chords. The posterior distribution median, 16th, and 84th percentiles are provided in Table \ref{tab:shape}\footnote{See Appendix \ref{cornerplots} for corner plots.}. 

\begin{deluxetable}{lcc}
\tablecaption{Weywot's Triaxial Shape\label{tab:shape}}
\tablewidth{0pt}
\tablehead{
\colhead{Parameter} & \colhead{} & \colhead{Posterior}
}
\startdata
Long axis (km)              & $a$          & $82^{+5}_{-3}$         \\
Intermediate axis (km)      & $b$          & $73^{+5}_{-3}$         \\
Polar axis (km)             & $c$          & $65^{+8}_{-12}$        \\
Pole R.A. (deg)             & $\alpha_p$   & $267.6^{+0.2}_{-0.2}$  \\
Pole dec. (deg)             & $\delta_p$   & $52.2^{+0.2}_{-0.2}$   \\
Lightcurve phase (deg)      & $\phi$       & $24^{+8}_{-8}$         \\
Ephemeris offset (km)       & $f$          & $-58^{+3}_{-3}$          \\
     & $g$          & $53^{+8}_{-9}$         \\
\hline
Volumetric radius (km)      & $r_{\rm eq}$ & $73^{+5}_{-5}$         \\
Polar axis ratio        & $c/a$        & $0.78^{+0.08}_{-0.13}$ \\
Intermediate axis ratio & $b/a$        & $0.88^{+0.04}_{-0.04}$ \\
\enddata
\tablecomments{Reported values show the median, 16th, and 84th percentile values. Pole coordinates are given at the time of the June 2023 occultation. The ephemeris offset is with respect to the \citet[][]{Proudfoot2025Weywot} ephemeris of Weywot (relative to Quaoar) and the NIMAv20 ephemeris of Quaoar (relative to the Solar System barycenter; \citealt[][]{Desmars2015}).}
\end{deluxetable}

\section{Discussion}
\label{WeywotDiscussion}

From the lightcurve and occultation analysis, we have obtained new information about Weywot's tidal state and 3D shape. Our assumption that Weywot's spin is synchronized with the mutual orbit period is supported by the observation that Weywot's lightcurve is nearly in-phase with its angular separation from Quaoar (Figure \ref{figure:m_r}). Weywot's lightcurve, which shows two cycles over the $\sim 12.427$-day rotation period with a peak-to-peak amplitude of $\Delta m = 0.119 \pm 0.036$ mag, indicates Weywot's shape is elongated, and provided the lightcurve phase Weywot showed during the June 2023 occultation \citep{Estela2026}. The occultation chords were refitted using the lightcurve measurements as constraints. Weywot's 3D shape was measured to be moderately prolate and significantly oblate, with an intermediate axis ratio $b/a = 0.88 \pm 0.04$ and polar axis ratio $c/a = 0.78^{+0.08}_{-0.13}$. The volume-equivalent radius was estimated to be $r_\text{eq} = 73 \pm 5$ km (Section \ref{occ}, Figure \ref{fig:shapemodel}, Table \ref{tab:shape}). 

Although no current dynamical measurements of Weywot's mass have been made, we can provide an upper bound on its mass using our shape model. Assuming Weywot has the same density as Quaoar \citep[$1751 \pm 13$ kg m$^{-3}$;][]{Proudfoot2025Weywot}, we derive an upper limit on Weywot's mass of $5.0 \times10^{18}$ kg, using the $3\sigma$ upper-limit to $r_\text{eq}$. Using a more realistic density for a body of Weywot's size, $\sim700$ kg m$^{-3}$, we estimate Weywot's mass to be $1.1\times10^{18}$ kg. This is very similar to the mass of Namaka, Haumea's smaller satellite \citep{Proudfoot2025haumea}. 

Such an estimate for the mass of Weywot is extremely useful to better understand the effect of Weywot on Quaoar's ring system. One possible explanation of the ring arcs present in both of Quaoar's rings is the dynamical effect of Weywot \citep{Rodriguez2023}, although doubt has been cast on this idea, as Weywot's eccentricity has been revised to much smaller values \citep{Proudfoot2025Weywot,BragaRibas2025}. A small mass for Weywot further weakens this hypothesis. \citet{Proudfoot2025Weywot,Proudfoot2025moonlet} speculated that Weywot could induce bending waves in the outer ring due to Weywot's inclination relative to the ring system ($\sim5\degr$). Our measurement of a relatively modest mass limits the strength of bending modes, although full simulations are needed to fully determine whether they are dynamically relevant. The recent discovery of a small satellite nearer the rings \citep{Nolthenius2025Moonlet} suggests that ring dynamics may instead be dominated by yet-to-be discovered ring-moons, rather than Weywot's influence.

The nominal solution for Weywot's shape is inconsistent with a fluid Roche equilibrium figure at Weywot's current semimajor axis. This implies Weywot’s interior resists self-gravity's attempt to make its shape that of a fluid Roche ellipsoid. Under the assumption that Weywot is a rubble-pile satellite potentially formed out of circum-Quaoar material, such that it has no tensile strength, internal friction between aggregates allows it to access a wider variety of triaxial ellipsoidal shapes. \cite{Sharma2009} developed equilibrium landscapes for triaxial rubble pile satellites which are constrained by empirical evidence that most gravel piles have an angle of repose $\phi_F \lesssim 40^{\circ}$. Weywot's triaxial shape at its current semimajor axis can be explained by a moderate internal friction angle of $5-10^{\circ}$, and thus is fully consistent with the rubble pile interpretation. The \citet{Sharma2009} model allows Weywot's triaxial shape to be in equilibrium down to a semimajor axis of $\simeq 2$ Quaoar-radii ($R_Q$). Therefore, if Weywot formed around Quaoar and is a rubble pile supported only by internal friction, it finished accreting material beyond $\sim 2 R_Q$. This falls just outside of Quaoar's current co-rotation distance.

The measured shape of Weywot and its nonzero orbital eccentricity imply Weywot's 1:1 spin-orbit resonance (SOR) may overlap with the 3:2 SOR \citep{Murray2000}. This allows the possibility that Weywot's spin axis may have been attitude unstable and/or chaotically tumbling at some point in the past \citep{Wisdon1987Tumble}. Our measurement of Weywot's rotational lightcurve, in particular the time-sampling of the the F350LP data over many rotations, implies a currently stable rotation axis. Similar to many satellites in the Solar System \citep{Thomas1998JupiterInner, Thomas2013SaturnInner}, it is possible that Weywot has undergone chaotic tumbling in the past before returning to principal axis rotation. 


Using Weywot's updated absolute magnitude and size, the $V$-band geometric albedo of Weywot ($p_V$) can be updated. The rotationally averaged absolute magnitude is $H_V = 8.484 \pm 0.042$ mag. The rotationally averaged, projected-area-equivalent radius during the observations is $\mathcal{R} = 71^{+6}_{-7}$ km. Using

\begin{equation}
    p_V = \frac{10^{-0.4(H_V - V_{\odot})}}{\mathcal{R}^2},
\end{equation}

\noindent where $\mathcal{R}$ is in units of au and $V_{\odot} = -26.74$ mag, we obtain $p_V = 0.036^{+0.010}_{-0.006}$.  

Weywot's low geometric albedo and measured phase curve reinforce the dichotomy between the surfaces of Quaoar and Weywot. Quaoar's surface \citep[$p_V = 0.124 \pm 0.006$;][]{Pereira2023} is $\approx 3.6 \times$ more reflective than Weywot's in $V$-band. Comparing to the broader population, Weywot's albedo falls at the lower end of the geometric albedo distribution of TNOs and Centaurs \citep{Muller2020}.  Weywot's low albedo and H$_2$O spectral type (A. C. Souza Feliciano, submitted) supports previous work that associated the H$_2$O spectral type with relatively darker surfaces \citep{discotnos}. While we leave the complete analysis of Quaoar's HST photometry to a future study, we measured Quaoar's phase curve using a similar methodology to be $\beta = 0.177 \pm 0.012$, shallower than Weywot's at the 3.4-$\sigma$ level. Extrapolating to larger phase angles points to Weywot's phase integral $q$ being low ($q \lesssim 0.5$), such that Weywot falls into the low-albedo, small phase integral class of small bodies, typical of cometary nuclei and C-type asteroids \citep{Verbiscer2022}. 

The albedo and composition of Weywot could be representative of Quaoar's rings. With a diameter of $D=146 \pm 11$ km, Weywot is expected to be undifferentiated, and internal heating through radioactive isotopes is inadequate to alter its bulk composition \citep[e.g.,][]{Bierson2019,Loveless2022}. We estimate that additional heating through tidal de-spinning raised Weywot's internal temperature by $\Delta T\ll1$ K. If Weywot formed from the same material as the rings, internal processes within Weywot are unlikely to alter the surface composition. Alternatively, Weywot may be the source of ring material \citep[e.g.,][]{Stern2006Impact, Porter2015Ejecta}. Impacts on Weywot that launch ejecta at escape velocities have likely populated the ring system to some degree. Future work could explore the expected contribution Weywot has made to the ring mass through impact ejecta.


In this way, unless the ring material is actively replenished by processes within Quaoar, ring particulates may also have a low albedo, and future spectroscopy of Weywot's surface could provide a proxy for ring composition. 

Weywot is among the smallest known satellites in trans-Neptunian space in terms of secondary-to-primary mass ratios ($q \lesssim 0.5$\%). In giant impact simulations with differentiated progenitors, \cite{Arakawa2019} found that satellites with $q \lesssim 0.5$\% form with ice mass fractions $f_{\rm ice} \sim 100$\%. This is in agreement with the bright water-ice dominated surfaces of Haumea's satellites and Pluto's minor satellites, all which have $p_V \gtrsim 0.5$ \citep{Fraser2009HaumeaSatellites, Cook2018Irregular, Estela2025Hiiaka}. Weywot and Dysnomia, on the other hand, have contrastingly dark surfaces ($p_V \lesssim 0.05$). This dichotomy in surface types may be the exterior representation of different bulk ice mass fractions ($f_{\rm ice} < 100$\%), consistent with Weywot and Dysnomia not being sourced from the icy mantle of a differentiated world as was suggested for Haumea's satellites and Pluto's minor satellites. Perhaps Dysnomia and/or Weywot formed from large undifferentiated planetoids. Moreover, \cite{Arakawa2019} simulated impacts between progenitors with mass fractions $q \geq 50 \%$ and the role of internal strength was not considered. Dysnomia and Weywot, conversely, have $q \lesssim 0.5$\% and have nonzero rigidity; Dysnomia and Weywot's current physical properties may represent their pre-impact properties, where they survived massive impacts with their hosts due to internal strength \cite[e.g.,][]{Denton2025Kiss}. Future work could simulate collisions that consider rigidity, and use progenitors that have much smaller mass ratios to see if these small satellites can survive massive impacts relatively undisturbed.

Evidenced by observations of their surface chemistry and/or tidal evolution modeling, past or present internal geologic activity and/or differentiation, typically involving subsurface liquid water reservoirs, has been proposed for a handful of TNOs \cite[e.g.,][]{Brown2007Haumea, Nimmo2016Reorientation, Nimmo2023, emery2024, Glein2024Moderate, Kiss2024Hotspot}. One of the open questions regarding dwarf-planets is the degree to which silicates in their interior were hydrated by liquid water \citep[e.g.][]{Dunham2019haumea, Farkas2022Serpentinization, Noviello2022Haumea, Farkas2023Serpentinization}. One potential avenue to sample the interiors of large TNOs is by observing the surface chemistry of their collisional remnants. If Weywot formed through a giant-impact, it remains plausible that it was sourced from material deep within a massive progenitor body. For this reason, future efforts should look to constrain the surface mineralogy of collisionally-produced satellites of the dwarf planet class TNOs.

\section{Summary}
11 orbits of Hubble Space Telescope data were used to constrain the size, 3D shape and geometric albedo of Quaoar's largest satellite Weywot. Weywot's independent photometric phase curve was measured for accurate estimation of its absolute magnitude. Weywot likely formed as the result of a massive collision in Quaoar's past, potentially forming from the same material making up the ring system. Alternatively, Weywot may be the source of ring material. A summary of the results and interpretations is as follows:

\begin{enumerate}
    \item The timescale for Weywot's spin to synchronize with its orbit as the result of tidal torques was estimated to be $<150$ Myr. Weywot's lightcurve also oscillates in-phase with its apparent angular separation from Quaoar, supporting synchronous rotation as well as a triaxial shape for Weywot.
    
    \item The lightcurve solution was used as priors to fit a stellar occultation by Weywot in June 2023 \citep{Estela2026}. Weywot's 3D shape was measured to be moderately prolate and significantly oblate, with an intermediate axis ratio $b/a = 0.88 \pm 0.04$ and polar axis ratio $c/a = 0.78^{+0.08}_{-0.13}$. The volume-equivalent radius was estimated to be $r_\text{eq} = 73 \pm 5$ km.  If Weywot formed around Quaoar and is a rubble pile supported only by internal friction, it finished accreting material beyond $\sim2$ Quaoar-radii. This falls just outside of Quaoar’s current co-rotation distance.

    \item An upper limit to Weywot's mass was derived to be $5.0 \times10^{18}$ kg, much lower than previous results. This reduces Weywot's influence on ring dynamics.
    
    \item The low geometric albedo of Weywot, $p_V = 0.036^{+0.010}_{-0.006}$, places Weywot's surface among the darkest known for TNOs. This suggests Weywot's ice-mass fraction $f_{\rm ice} < 100$\%, and that Weywot is not sourced from the water-ice mantle of a differentiated pre-impact progenitor as suggested for Pluto's minor satellites and Haumea's satellites. Weywot may have formed from the debris of an undifferentiated dwarf-planet progenitor, or is a surviving collisional fragment of a smaller TNO whose surface was depleted in volatiles. On the other hand, perhaps Weywot is sourced from the deep water-ice depleted interior of a fully differentiated, collisionally disrupted world. 
\end{enumerate}

\begin{acknowledgments}
This research is based on observations made with the NASA/ESA Hubble Space Telescope obtained from the Space Telescope Science Institute, which is operated by the Association of Universities for Research in Astronomy, Inc., under NASA contract NAS 5–26555. These observations are associated with programs 17417 and 17827. The HST data presented in this article were obtained from the Mikulski Archive for Space Telescopes (MAST) at the Space Telescope Science Institute. The specific observations analyzed can be accessed via \dataset[doi: 10.17909/c1xh-2f85]{https://doi.org/10.17909/c1xh-2f85}.

C.C. acknowledges the support of the American Astronomical Society and the National Science Foundation in the form of an International Travel Grant. B.P. acknowledges the support of the University of Central Florida Preeminent Postdoctoral Program (P$^3$). 

\end{acknowledgments}

\begin{contribution}
C.C. and B.P. lead the overall analysis in the paper and consulted with E.F.V. and B.J.H.. E.F.V. was the P.I. of HST program 17827 and led the occultation campaign. C.C. conducted photometry on the HST data, and consulted with B.J.H. and J.L.O.. C.C. and B.P. developed the photometric and tidal evolution models. B.P. conducted the occultation analysis. I.S. contributed to the rubble-pile Roche-equilibrium figure interpretation of Weywot's shape. 

\end{contribution}

\clearpage
\appendix 
\twocolumngrid
\section{Astrometric Measurements}
In Table \ref{astrometry} we tabulate the computed relative astrometric measurements from the F606W images. The values correspond to Weywot's position on-sky relative to Quaoar.

\begin{deluxetable}{ccccccc}
\tablecaption{Astrometric Measurements from F606W Images\label{astrometry}} 
\tablehead{Date & $\Delta x \hspace{0.01cm} \text{cos}(\delta)$ & $\sigma_{\Delta x \hspace{0.01cm} \text{cos}(\delta)}$ & $\Delta y$ & $\sigma_{\Delta y}$ & $r$ & $\sigma_r$ \\ (Geocentric MJD) & ($''$)& ($''$)& ($''$)& ($''$)& ($''$)& ($''$)}
\startdata
60856.78121 & 0.41857 & 0.00068 & 0.10716 & 0.00117 & 0.43207 & 0.00072   \\
60858.34953 & 0.19170 & 0.00071 & 0.17938 & 0.00257 & 0.26254 & 0.00183   \\
60860.24879 & -0.21583 & 0.00204 & 0.12682 & 0.00138 & 0.25033 & 0.00189  \\
60861.75842 & -0.41903 & 0.00151 & -0.00079 & 0.00054 & 0.41903 & 0.00151 \\
60863.32680 & -0.38905 & 0.00065 & -0.12584 & 0.00068 & 0.40890 & 0.00065 \\
60865.15079 & -0.06102 & 0.00565 & -0.18621 & 0.00514 & 0.19595 & 0.00519 \\
\enddata
\tablecomments{Each MJD date is the mid-time of all exposures in that orbit, not corrected for light travel time. $\Delta x$ corresponds to the differece in right ascension and $\Delta y$ corresponds to the change in declination ($\delta$).}
\end{deluxetable}

\section{Estimating the Synodic Offset}
\label{synodic}
We estimate Weywot's synodic rotation period as follows. Define a non-rotating, orthogonal basis on Quaoar's center of mass using the orbit pole as one basis vector. Using barycentric Cartesian coordinates for the Earth and Quaoar at the observing times, convert the Earth-Quaoar position vector to rotations around the orbit pole. Defining $\delta \vartheta$ as the small rotation from time $t_{i} \to t_{i+1}$, with $\delta t = t_{i+1} - t_{i}$, the perceived synodic spin period at $t_{i+1}$ is 

\begin{equation}
    P_{\text{syn}} = P_{\text{sidereal}} - 2\pi \left(\frac{\delta \vartheta}{\delta t} \right)^{-1}.
\end{equation}

\noindent We estimate a range of synodic offsets of $2\pi (\delta \vartheta / \delta t)^{-1} \in [-3.7, 7.4]$ min, with an average of $-3.2$ min. This offset has a negligible impact on our results. 

\clearpage
\section{Corner Plots}
\label{cornerplots}
\begin{figure*}
\centering
 	\includegraphics[scale = 0.7]{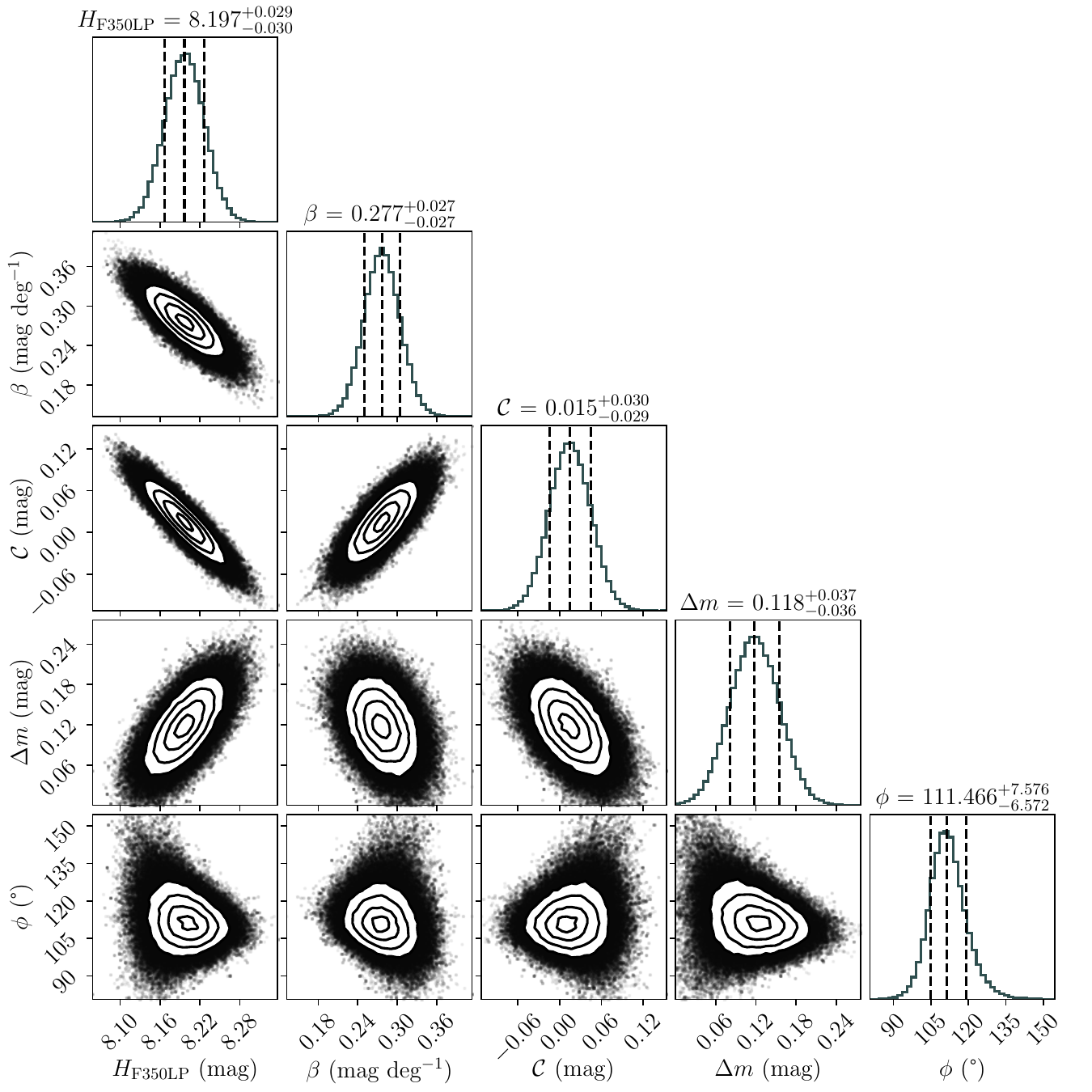}
	\caption{
    A corner plot showing the posterior distribution solution for the model of Weywot's photometry. Contours in the 2D histogram represent the 11.8\%, 39.3\%, 67.5\%, and 86.4\% confidence levels and black points correspond to individual samples from the MCMC chain. Dashed lines in each parameter's 1D histograms represent the $1\sigma$ confidence interval and the median. See Sections \ref{lcmodel} \& \ref{lcresults} for further details.
 \label{figure:Weywot_corner}
 }
\end{figure*}

\begin{figure*}
    \centering
    \includegraphics[width=\textwidth]{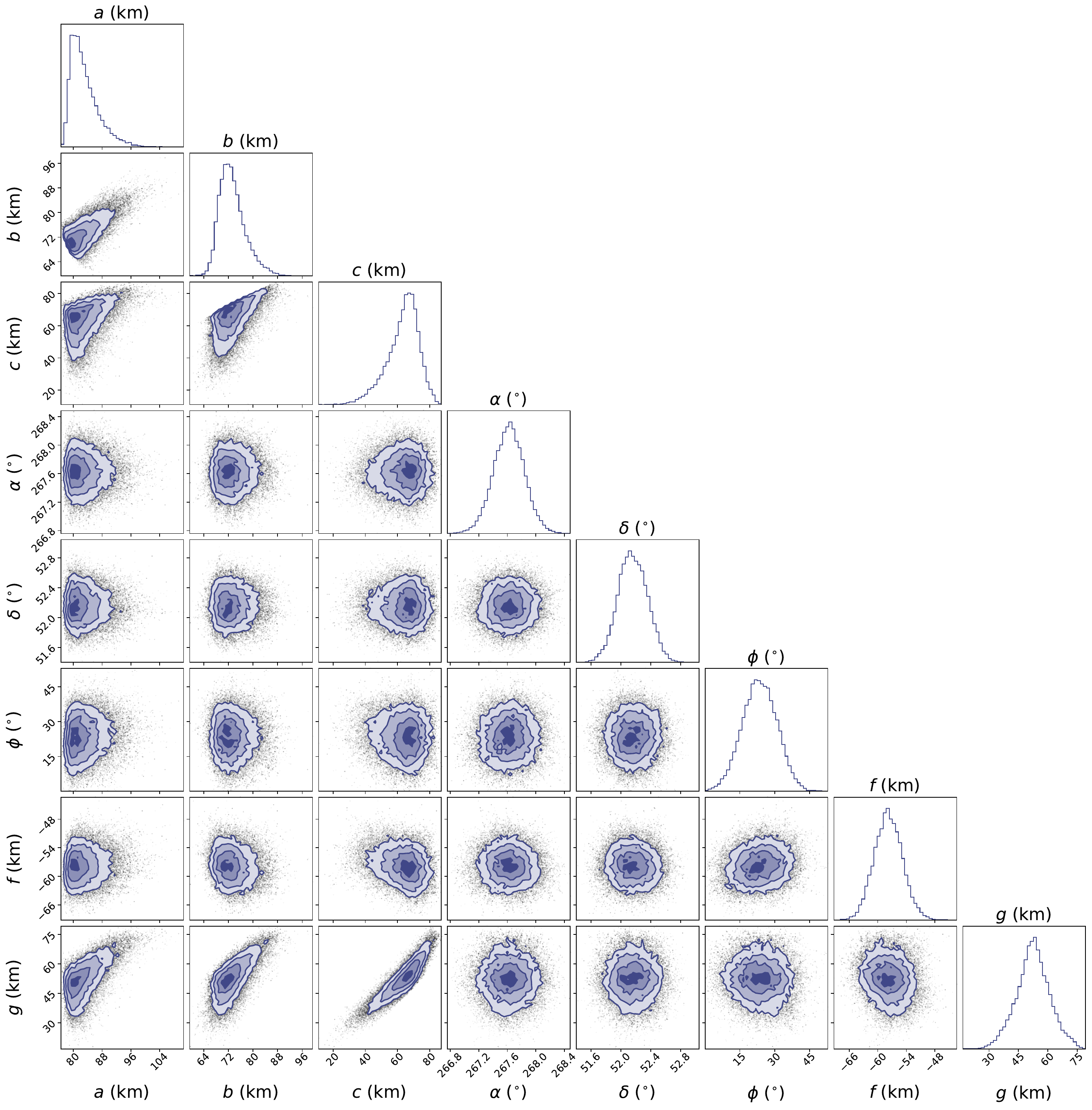}
    \caption{A corner plot showing the posterior distribution of the ellipse fit. Contours in the 2D histogram represent the 11.8\%, 39.3\%, 67.5\%, and 86.4\% confidence levels and black points correspond to individual samples from the MCMC chain. The 1D histograms show the posterior distributions for each parameter in the ellipse fit model. See Section \ref{occ} for further details.}
    \label{fig:corner_shape}
\end{figure*}

In this section we show corner plots for the MCMC tests noted in the main text. The corner plot corresponding to Weywot's photometric and lightcurve parameters (Section \ref{lcresults}, Table \ref{W_lcparams}) are shown in Figure \ref{figure:Weywot_corner}. Figure \ref{fig:corner_shape} shows the corner plot corresponding to Weywot's shape model (Section \ref{occ}, Table \ref{tab:shape}).


\section{Filter Conversion: Hubble F606W to Johnson V}
\label{filterconversion}
Converting from F606W to $V$-band requires knowledge of the reflectance spectrum of the object and absolute flux calibration. The $H_{F606W}$ magnitudes in the main text are in the VEGAMAG system. Converting back to an instrumental magnitude requires removing the VEGAMAG zeropoint in the F606W filter, where $Z_{F606W}$ = 29.995 mag. Fully calibrated, flat-fielded, and charge transfer efficiency-corrected frames provided by MAST provides the conversion factor $\xi$ to convert a measured count rate in e$^{-}$ s$^{-1}$ to flux density units \citep{Bohlin2014}. The mean and standard deviation of $\xi$ from the F606W images was (119308.4 $\pm$ 2.0) $\times 10^{-24}$ erg (e$^{-1}$)$^{-1}$ cm$^{-2}$ \r{A}$^{-1}$. The mean flux density in F606W corresponding to the absolute magnitude is

\begin{equation}
    \langle f(\lambda_{F606W}) \rangle = \xi 10^{-0.4(H-Z)}.
\end{equation}

\noindent Magnitudes on the Johnson system can be calculated knowing the mean flux density of the object in $V$-band ($\langle f(\lambda_{V})\rangle$) using

\begin{equation}
    m_V = -2.5\text{log}_{10}\left( \frac{\langle f(\lambda_{V})\rangle}{\langle f(\lambda_{V,0}\rangle}  \right),
\end{equation}

\noindent where $\langle f(\lambda_{V,0})\rangle = 3.62701\times 10^{–9}$ erg s$^{-1}$ cm$^{-2}$ \r{A}$^{-1}$ is the zeropoint flux \citep{Willmer2018solarmag}. $\langle f(\lambda_{F606W}) \rangle$ can be converted to $\langle f(\lambda_{V})\rangle$ as follows. Let $f(\lambda)$ be the reflected flux density of the TNO with the solar component preserved, and $S_x(\lambda)$ be the energy response of filter $x$. The mean flux density expected in filter $x$ is 

\begin{equation}
    \langle f(\lambda_x) \rangle = \frac{ \int f(\lambda) S_x(\lambda) d\lambda}{\int S_x(\lambda) d\lambda},
\end{equation}

\noindent granting measurement of the flux ratio

\begin{equation}
    r = \frac{\langle f(\lambda_{F606W}) \rangle}{\langle f(\lambda_{V}) \rangle}
\end{equation}

\noindent such that $\langle f(\lambda_{V}) \rangle = r \hspace{0.05cm} \langle f(\lambda_{F606W}) \rangle$.

\clearpage
\bibliography{zzrefs}{}
\bibliographystyle{aasjournalv7}

\end{document}